\def\year{2021}\relax
\documentclass[letterpaper]{article} 
\usepackage{aaai21}  
\usepackage{times}  
\usepackage{helvet} 
\usepackage{courier}  
\usepackage[hyphens]{url}  
\usepackage{graphicx} 
\usepackage{natbib}  
\usepackage{caption} 

\usepackage{algorithm} 
\usepackage{algpseudocode}

\title{Validation of Deep Convolutional Generative Adversarial Networks for \\ High Energy Physics Calorimeter Simulations}

\author {
    Florian Rehm\textsuperscript{\rm 1,2}, 
    Sofia Vallecorsa\textsuperscript{\rm 1},
    Kerstin Borras\textsuperscript{\rm 2,3},
    Dirk Krücker\textsuperscript{\rm 3}\\
}
\affiliations {
    \textsuperscript{\rm 1} CERN, Switzerland \\
    \textsuperscript{\rm 2} RWTH Aachen Universtiy, Germany \\
    \textsuperscript{\rm 2} DESY, Germany \\
    florian.matthias.rehm@cern.ch
}

\begin{document}

\maketitle


\begin{abstract}
In particle physics the simulation of particle transport through detectors requires an enormous amount of computational resources, utilizing more than 50\% of the resources of the CERN Worldwide Large Hadron Collider Grid.
This challenge has motivated the investigation of different, faster approaches for replacing the standard Monte Carlo simulations.
Deep Learning Generative Adversarial Networks are among the most promising alternatives. Previous studies showed that they achieve the necessary level of accuracy while decreasing the simulation time by orders of magnitudes. 
In this paper we present a newly developed neural network architecture which reproduces a three-dimensional problem employing 2D convolutional layers and we compare its performance with an earlier architecture consisting of 3D convolutional layers.
The performance evaluation relies on direct comparison to Monte Carlo simulations, in terms of different physics quantities usually employed to quantify the detector response. We prove that our new neural network architecture reaches a higher level of accuracy with respect to the 3D convolutional GAN while reducing the necessary computational resources.
Calorimeters are among the most expensive detectors in terms of simulation time. Therefore we focus our study on an electromagnetic calorimeter prototype with a regular highly granular geometry, as an example of future calorimeters.
\end{abstract}

\section{Introduction}
\label{sec:introduction}
\noindent 
Today, particle detectors are simulated using Monte Carlo-based methods such as the Geant4 toolkit \cite{Geant4}. 
Due to complex detector geometries and manifold diverse physical processes, Monte Carlo simulations have a high computational cost and require, currently, more than half of the Worldwide Large Hadron Collider (LHC) Grid resources \cite{RoadmapHEP}. 
In the future High Luminosity phase of the LHC (HL-LHC) around 100 times more simulated data are expected \cite{HL-LHC}. Even when taking the technological improvements into account, this will exceed considerably the expected computational resources at the HL-LHC regime. 

In order to overcome this obstacle, intense research is ongoing for developing faster alternatives to the standard Monte Carlo simulation.
The main difficulty in searching for a faster simulation approach is given by the high level of accuracy required.
Based on deep generative models, several prototypes demonstrated an enormous potential by achieving similar accuracy than the traditional simulation \cite{de2017learning,paganini2017calogan,Salamani2018,dijet,gan_lhcb}. Generative Adversarial Networks (GANs) represent an example of such models: so far, they have been employed primarily for simulating calorimeters \cite{EnergyGAN}. 
In this paper we describe a novel GAN architecture which uses two-dimensional convolutional layers (Conv2D) for generating the
three-dimensional volume of an electromagnetic calorimeter. We compare our architecture to a previous three-dimensional convolutional (Conv3D) GAN model developed for the same detector use case \cite{EnergyGAN}. Results are compared to the corresponding Monte Carlo simulation in terms of multiple physics validation metrics. The Conv2D model improves the physics accuracy while reducing the computation time.

In the following we provide a review of related work, followed by a brief description of the electromagnetic calorimeter and the training data set. 
Next, we introduce our Conv2D prototype and we evaluate its performance in terms of physics accuracy and computation time. The last section summarizes our conclusions and plan for future work.

\section{Related Work} \label{sec:RelatedWork}
\noindent 
GANs belong to the group of generative models and are nowadays used for a large variety of different tasks \cite{GANS_Applications}. The whole GAN model consists of two deep neural networks: a generator network, which generates images from a latent vector consisting of random numbers, and a discriminator, trained to evaluate and distinguish between the generated images of the generator and the training or validation images. 
GANs are among the most promising approaches, which are being successfully investigated for replacing traditional Monte Carlo for simulating detectors and, in particular, calorimeters. Several studies demonstrate that GANs achieve similar levels of accuracy as Monte Carlo, while considerably decreasing the simulation time \cite{EnergyGAN} \cite{GAN_Gulrukh} \cite{WGAN_Thorben}.

\begin{figure*}[ht!]  
    \centering
    \includegraphics[width=.97\textwidth, clip=true]{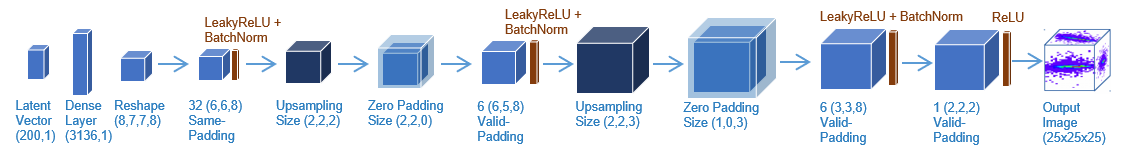}
    \caption{Convolutional 3D generator neural network architecture from \cite{EnergyGAN}. Input is a latent vector and output is the 3-dimensional shower image.}
    \label{fig:Conv3D}
\end{figure*}

In this research we compare our results with the Conv3D GAN model \cite{EnergyGAN} developed for the same calorimeter prototype. The model is available on GitHub \cite{GAN_Github} and its generator architecture is shown in figure \ref{fig:Conv3D} for reference. 
This model achieved in \cite{EnergyGAN} a large speed up (4 ms simulation time for a single electron compared to 17 s simulation time using Geant4 on a Intel Xeon processor) while reaching a good agreement with Monte Carlo in terms of physics \cite{EnergyGAN}.


\cite{Torben_Thesis} used a different Conv2D network architecture for simulating three dimensional calorimeter shower images. This was achieved by generating 12 two dimensional 15x7 images and stacking them to 3D images together. In our studies we have tested a similar architecture, based on stacking a set of two dimensional images, each representing one of the calorimeter layers. However this approach resulted in a worse performance, in terms of physics and speed up, with respect to the model we suggest in this research.

\section{Electromagnetic Calorimeters} \label{sec:Calorimeters}
\noindent 
By measuring particles energy, calorimeters are one of the core components of High Energy Physics experiments \cite{ECAL}. The primary particles enter and interact with the material of the calorimeter and deposit their energies by creating secondary particles as they pass through the detector. The secondary particles, in turn, create other particles by the same mechanisms leading to the generation of showers. While the shower evolves, the energy of the particles gradually reduces as it is absorbed or measured by the calorimeter sensors. 
In our work, we simulate electromagnetic calorimeters, which are specially designed to measure energies of electrons, positrons and photons that interact in the detector volume via electromagnetic interactions (mainly bremsstrahlung for electrons and pair production for photons).
As stated above, calorimeters represent computationally the most demanding simulations and are, therefore, the starting point for accelerating the simulation process. By representing each of the calorimeter sensors as a pixel in an image and the corresponding energy measurement as the pixel intensity, it is possible to interpret the detector output as an image and apply to it computer vision strategies and methods. It should be noted, however, that such "shower images" exhibit important differences with respect of typical RGB pictures, notably their high level of sparsity and large pixel intensity dynamic range, which can span several orders of magnitude.

\begin{figure}[]  
    \centering
    \includegraphics[width=.46\textwidth]{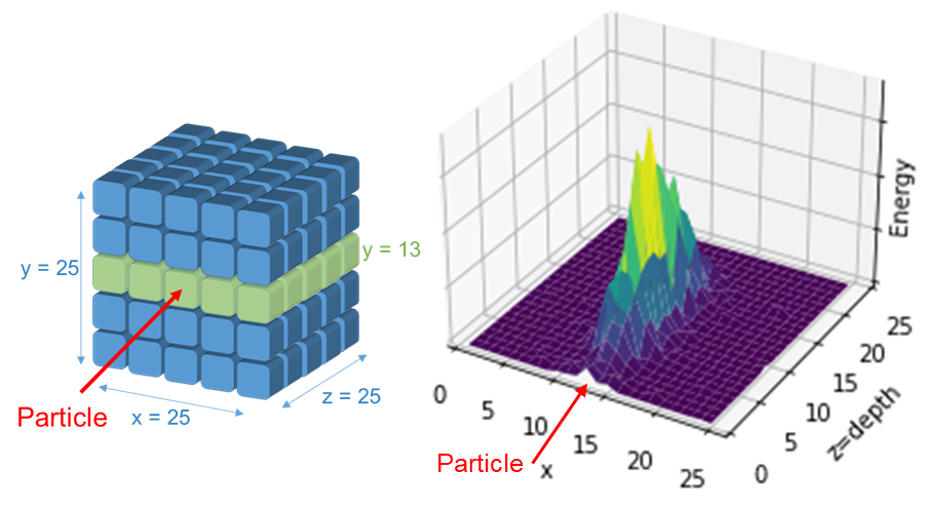}
    \caption{Representation of the 3D space (left) and an example  shower development at $y=13$ (right).}
    \label{fig:example_shower}
\end{figure}

Our training and test data sets contain particle shower images recorded by the calorimeter, simulated using Geant4 for a future electromagnetic calorimeter prototype. For our study, we use $200\,000$ electron shower images with initial energies in the range of 2-500 GeV and a 3-dimensional shape of $\mathrm{25x25x25}$ pixels. Details on the calorimeter architecture and the data set can be found in \cite{EnergyGAN}. 
Figure \ref{fig:example_shower} shows an example shower image cutout in the central layer ($y=13$), where the particle enters the calorimeter perpendicular to the $x$-$y$-plane at position $x=13$, $y=13$ and $z=0$, the shower is evolving in the $z$-direction. 
Particle shower images are parameterised according to the energy of the primary particle $\mathrm{E_p}$, which is the energy of the electron entering the calorimeter volume and the total energy measured by the calorimeter, which we indicate as $\mathrm{Ecal}$ (for Electromagnetic CALorimeter). The relation between $\mathrm{Ecal}$ and  $\mathrm{E_p}$  depends on the type, geometry and material of the calorimeter. 

\section{3D Generative Adversarial Networks} 
\label{sec:GAN}

\begin{figure*}[ht!]  
    \centering
    \includegraphics[width=.96\textwidth, clip=true]{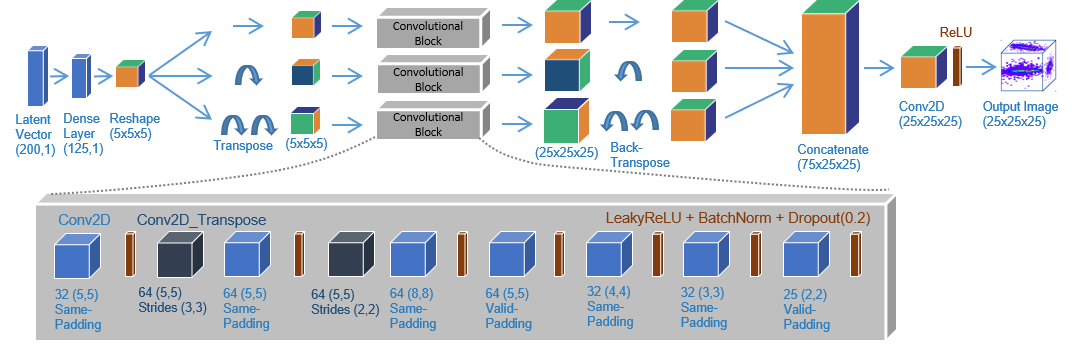}
    \caption{Conv2D Generator neural network architecture. Input is a latent vector and output is the 3-dimensional shower image.}
    \label{fig:generator}
\end{figure*}

\noindent 
Replacing Conv3D layers, in neural networks, with Conv2D layers reduces in general the computational complexity and therefore, computing resources and training time per epoch. However a non-trivial strategy is necessary to combine Conv2D layer in order to reproduce correctly a three-dimensional volume.
Our Conv2D generator architecture, shown in figure \ref{fig:generator}, uses a mix of 2D convolutional layers and transposed 2D convolutional (Conv2D\_transpose) layers (to increase the image size to the desired dimension). Additionally, we use batch normalization (BatchNorm) \cite{batchnorm}, a mix of rectified linear units (ReLU) and leaky rectified linear units activation functions (LeakyReLU) \cite{LReLU}, in order to account for the image sparsity, and dropout layers (Dropout) \cite{Dropout} to prevent mode collapse. The convolutional layers are organised in three branches in order to enable the model to correctly learn pixel-level correlations along the three canonical image dimensions.
Three-dimensional images, with $\mathrm{25x25x25}$ shapes are generated and parameterised (conditioned) using the input particle energy $E_p$. 
The hyper parameters, learning rate and learning rate decay, which we use for our model are found through a hyperparameter search with the Optuna tool \cite{optuna}.

Comparing the total number of parameters for the Conv3D and the Conv2D generator (table \ref{tab:models}), it can be noted that the Conv2D model has more than twice the number of parameters as the Conv3D model. Moreover, table \ref{tab:models} shows that the Conv3D network has most of its parameters (630k) in the first dense layer. This choice could represent a limitation, of the Conv3D model, in terms of representational power. On the other hand, the Conv2D model contains only a small fraction of its parameters in the dense layer. The advantage of having more parameters in the convolutional layers (meaning more convolutional layers, more channels and/or bigger filter sizes) is that convolutional layers perform better in learning images correlations. For this reason, we expect the Conv2D model to have a higher learning capability and to be able to solve more sophisticated tasks.

The architecture differences have a direct impact on the training and inference time. We bench-marked the inference process by running 20 warm-up steps and then 100 inference steps including 20 batches in each step. As shown in table \ref{tab:models}, the Conv2D inference time is much shorter, despite the larger number of parameters. The Conv2D model is also faster in terms of training time: with only 105s per epoch, instead of 291s for the Conv3D.
All GAN runs are performed on a NVIDIA Tesla V100-SXM2 GPU using TensorFlow version 2.0, CUDA version 10.2 and a batch size of 128. Unfortunately, Geant4 does not fully support GPU architectures at this time, its simulation time is measured on a Intel Xeon processor in \cite{EnergyGAN}. 

\begin{table}[h!]
 \centering
 \caption{Total number of parameters of the Conv2D and Conv3D generator networks and the number of parameters within the first dense layer. Inference times on GPU and the respective speed up compared to Monte Carlo.}
 \begin{tabular}{|c c c c c|} 
 \hline
 Model & Total & Dense & Inference & Speed Up\\ [0.5ex]
 
 \hline
 Conv2D & 2.05M   &	0.12M    & 5.4 s & $8\,000$x\\ 
  \hline
 Conv3D & 0.87M    &	0.63M &7.0 s & $6\,200$x    \\ 
 \hline
\end{tabular}
 \label{tab:models}
\end{table}

\section{Evaluation} \label{sec:Evaluation}
In this section we evaluate the performance of our new Conv2D model against the Conv3D model in \cite{EnergyGAN} and the Monte Carlo approach (Geant4).

Performance evaluation of generative models remains a difficult task for which, depending on the specific applications, several methods have been proposed \cite{GANevaluation}.
For evaluating the physics we measure the energy patterns across the calorimeter volume focusing on specific quantities that are typically used to characterize calorimeter properties.    

\begin{figure*}[ht!]  
    \centering
    \includegraphics[width=.8\textwidth, clip=true]{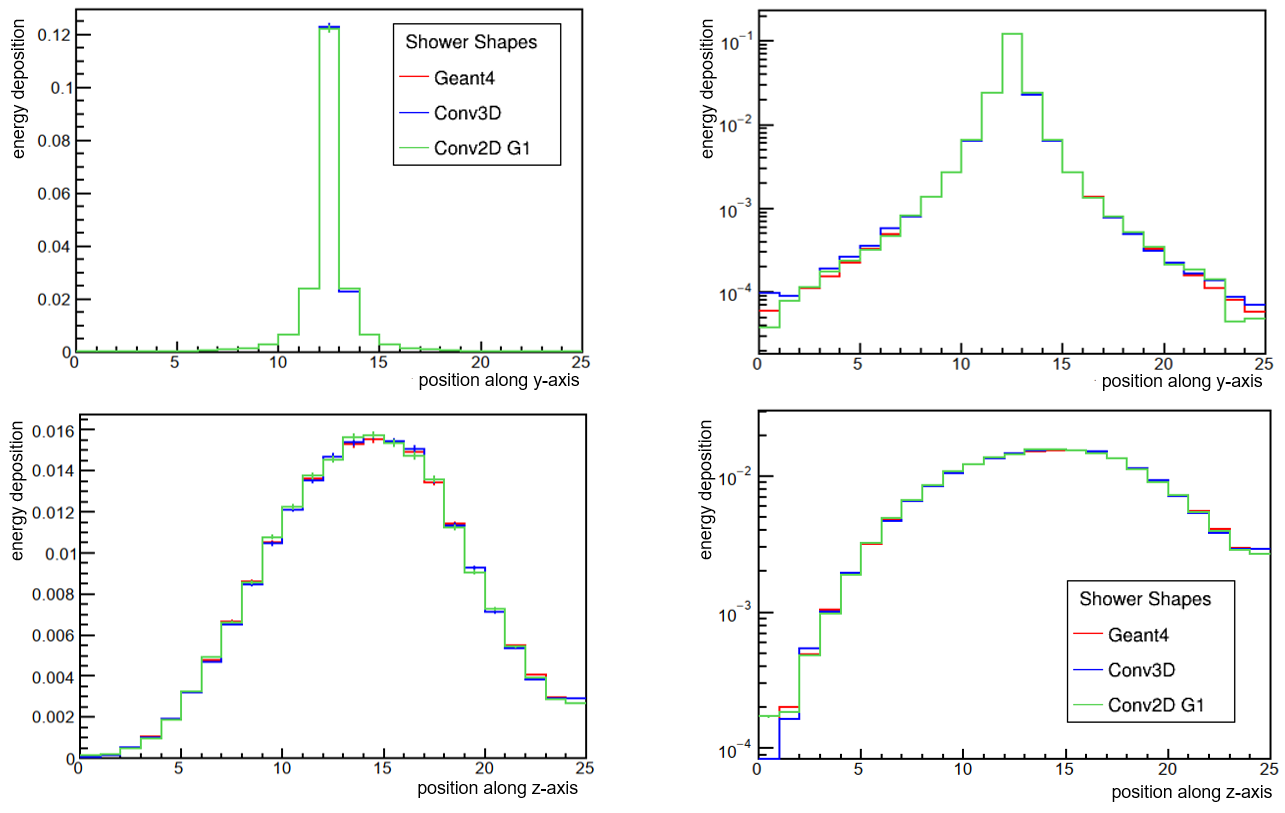}
    \caption{Particle shower distribution along the $y$- and $z$-axes on a linear energy axis scale (left) and on a logarithmic scale (right).}
    \label{fig:shower_shape}
\end{figure*}

Firstly, we build the particle shower shapes along the $y$- and $z$-axes for $20\,000$ samples within an energy range of 2-500 GeV. The shower development along the $x$-axis is similar to the $y$-axis and is therefore not shown. These shower shapes are shown in figure \ref{fig:shower_shape} (left) on a linear energy axis scale and (right) on a logarithmic scale. The GAN distributions agree with the expected Monte Carlo values within the uncertainties. In particular, previous studies had indicated poor performance at the edges of the simulated volume (the first and last cells along the $x$- and $y$-axes) \cite{GAN_Gulrukh}. In figure \ref{fig:shower_shape} we can see that both GAN models perform worse when the energy depositions become very small at the edges of the sensitive volume. However, the overall Kolmogorov-Smirnov (two-samples) test \cite{KS-Test}, yields probability above 80\% for most distributions and primary particle energies.

In order to simplify the evaluation during training and hyper-parameter tuning, we build a Mean Squared Error ($\mathrm{MSE}$) based on the above 2D shower shapes comparing the GAN distribution with Monte Carlo. We measured for the Conv2D model a $\mathrm{MSE}$ of 0.044 and for the Conv3D model 0.065, indicating a better accuracy (the lower the $\mathrm{MSE}$ the better), in agreement with the results from the shower shape plots from figure \ref{fig:shower_shape}.

\begin{figure}[ht!]  
    \centering
    \includegraphics[width=.45\textwidth, clip=true]{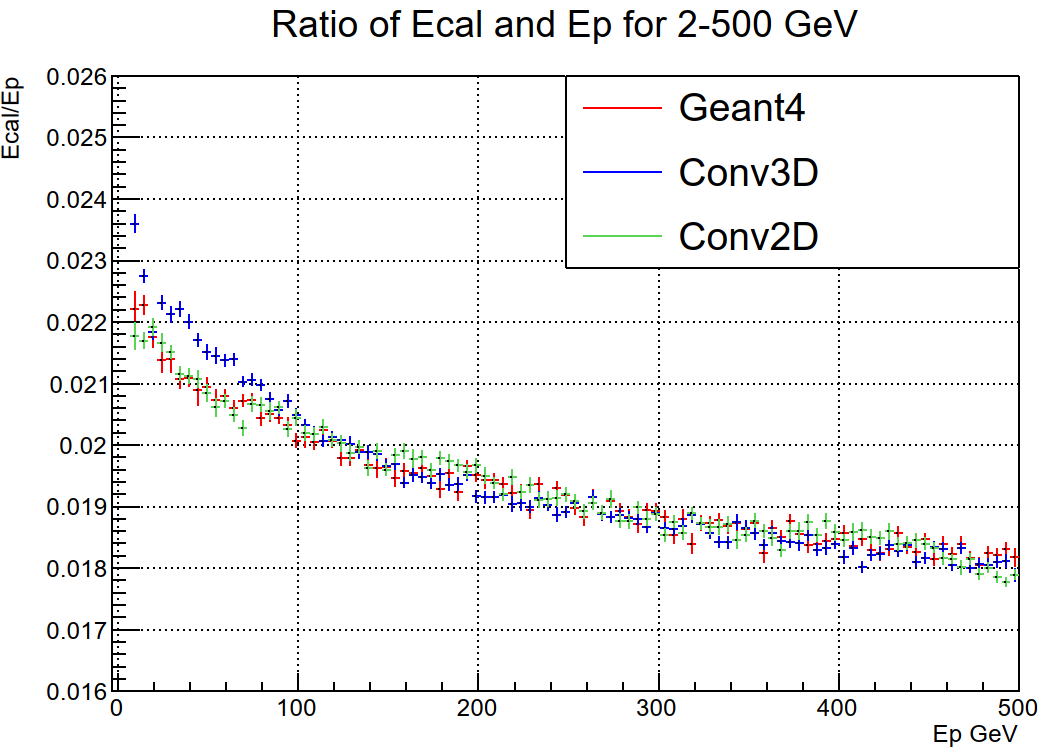}
    \caption{Sampling fractions comparing the Conv3D and Conv2D response to Geant4 over the full energy range.}
    \label{fig:samplingfraction}
\end{figure}

The sampling fraction is shown in figure \ref{fig:samplingfraction} as the ratio between the total measured energy $\mathrm{Ecal}$ and the initial particle energy $\mathrm{E_p}$. It represents the energy response of the generator network over the full energy range.
We can see that the Conv2D model maintains a exact agreement to Geant4 over the whole energy range, whereas the Conv3D model performs worse for $\mathrm{E_p}$ energies below 100 GeV.

\begin{figure}[ht!]  
    \centering
    \includegraphics[width=.42\textwidth, clip=true]{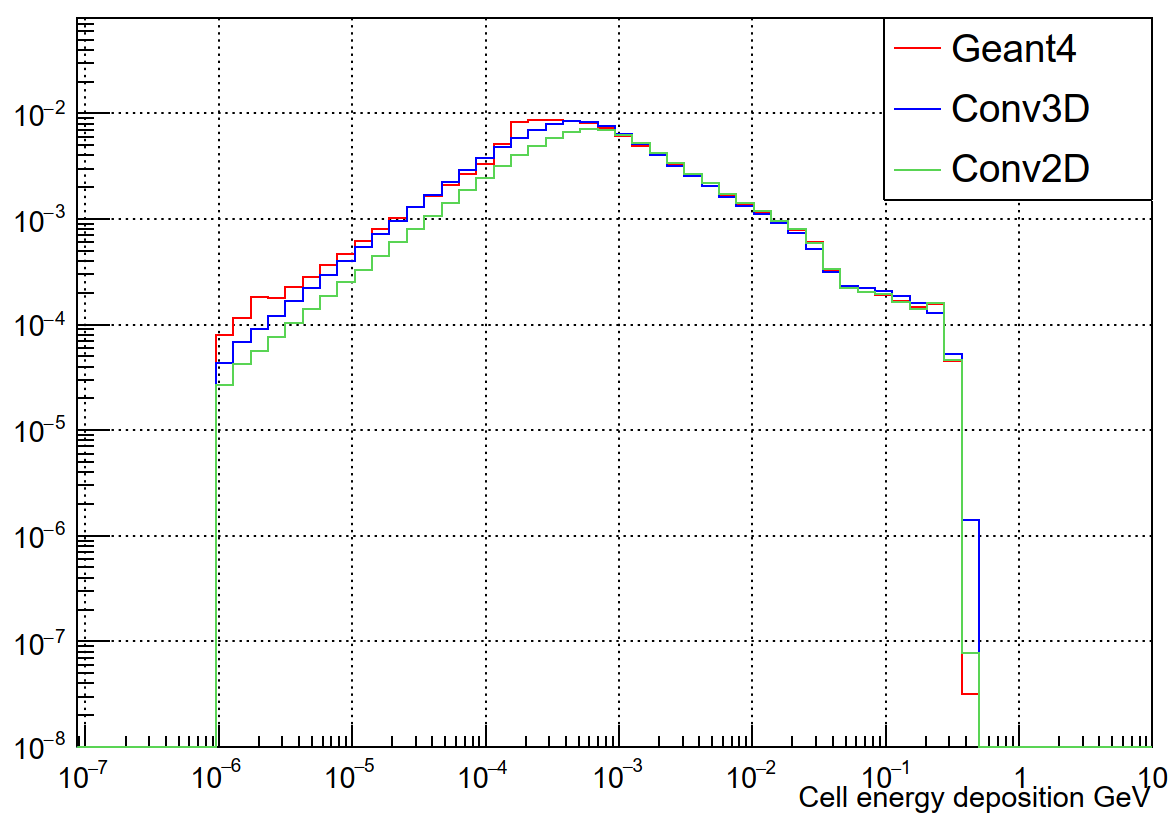}
    \caption{Histogram of cell energy deposition.}
    \label{fig:flatecal}
\end{figure}

For calorimeter simulations it is important that the single cell energy response is correctly reproduced by the GAN models. Figure \ref{fig:flatecal} shows the single cell energy deposition for $20\,000$ showers with input energies in the range of 2-500 GeV. The GAN description is not exact, in particular for small cell energy depositions, below 1 MeV. Additional research is required in order to improve the agreement. However, it should be noted, that we applied a minimal energy threshold of $10^{-6}$ GeV to take the detector energy resolution into account.

\begin{figure}[ht!]  
    \centering
    \includegraphics[width=.45\textwidth, clip=true]{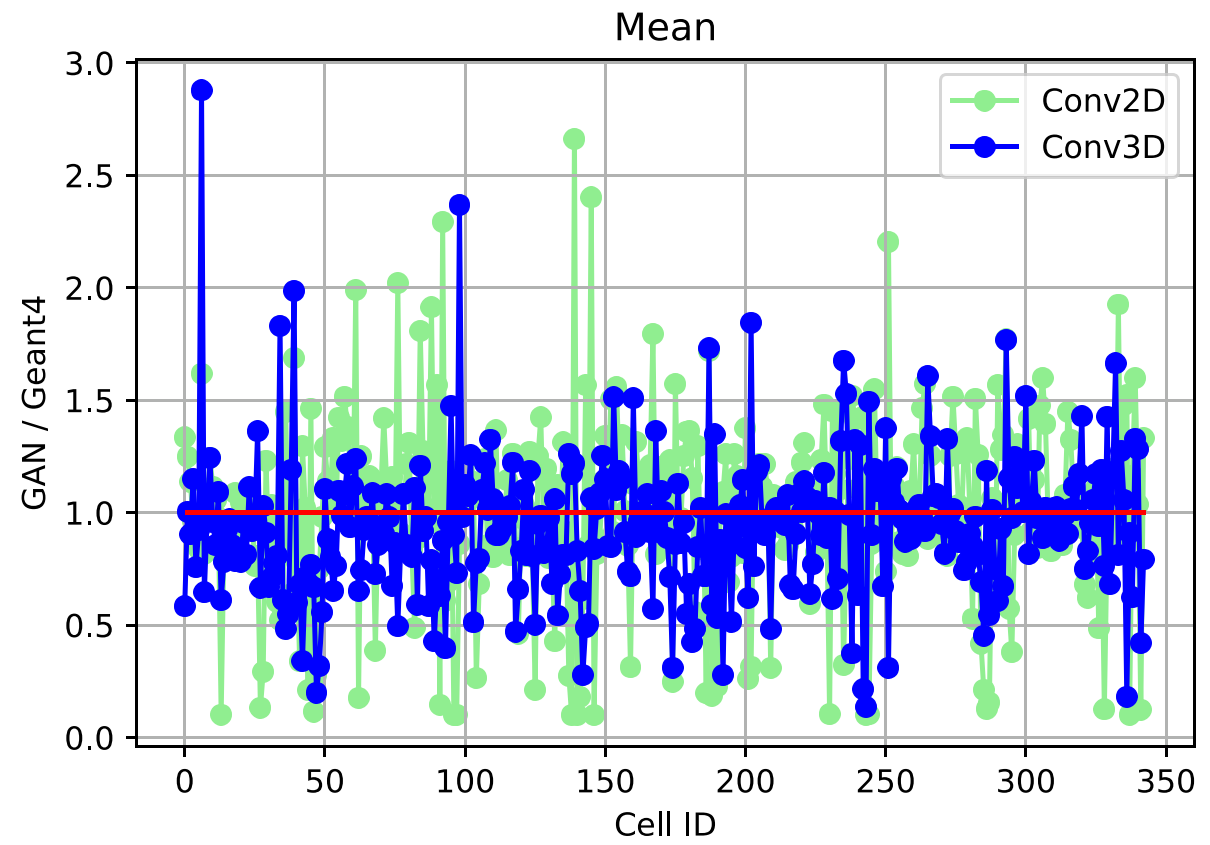}
    \caption{Single cell (pixel) energy distribution mean. The Conv2D model has an overall mean of $1.00$ and the Conv3D model an overall mean of $0.93$.}
    \label{fig:singlepixel_mean}
\end{figure}

\begin{figure}[ht!]  
    \centering
    \includegraphics[width=.45\textwidth, clip=true]{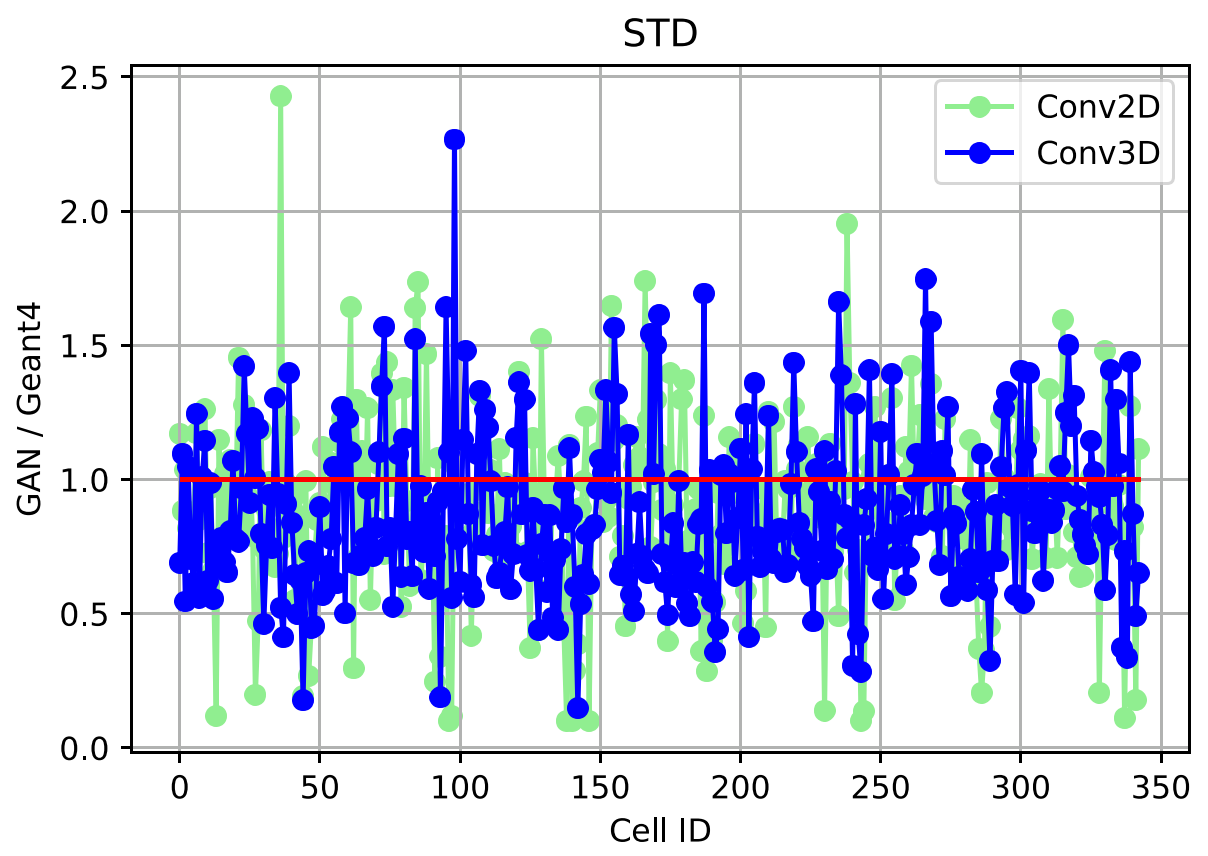}
    \caption{Single cell (pixel) energy standard deviations. The Conv2D model has an overall standard deviation of $0.90$ and the Conv3D model an overall standard deviation of $0.86$.}
    \label{fig:singlepixel_std}
\end{figure}

Figures \ref{fig:singlepixel_mean} and \ref{fig:singlepixel_std} compare the mean and standard deviations of single cell energy distributions in GAN and Monte Carlo samples for $1\,580$ showers with input energy $\mathrm{E_p}=250\pm2 \,$ GeV. For simplicity, we focus on the cells that constitute the core of the energy showers, where the larger energy depositions occur, and we analyse the cells with coordinates $x,y,z \in [10,16]$, for a total of $7\cdot7\cdot7=343$ cells. 
The expected ratio $=1$ is drawn as a red line in both figures, for comparison.

Figure \ref{fig:singlepixel_mean} shows that the mean energy deposition per cell is extremely close to Geant4 for the Conv2D model, their ratio averaging at $1.00$. On the other hand, the Conv3D model slightly underestimates the mean cell energy deposition with respect to Geant4, the average ratio close to $0.93$.
Similarly, figure \ref{fig:singlepixel_std} shows the GAN/Geant4 ratios in terms of the single cell energy distribution standard deviations (STD), which we use as a proxy for the single cell energy resolution. While both models underestimate this quantity, the Conv2D architecture exhibits values that are closer to the Monte Carlo prediction, with average ratios around $0.90$ against the $0.86$ obtained for Conv3D. In addition it is worthwhile for a further study to look also in to the tails of the shower to demonstrate the level of agreement.

\begin{figure}[ht!]  
    \centering
    \includegraphics[width=.45\textwidth, clip=true]{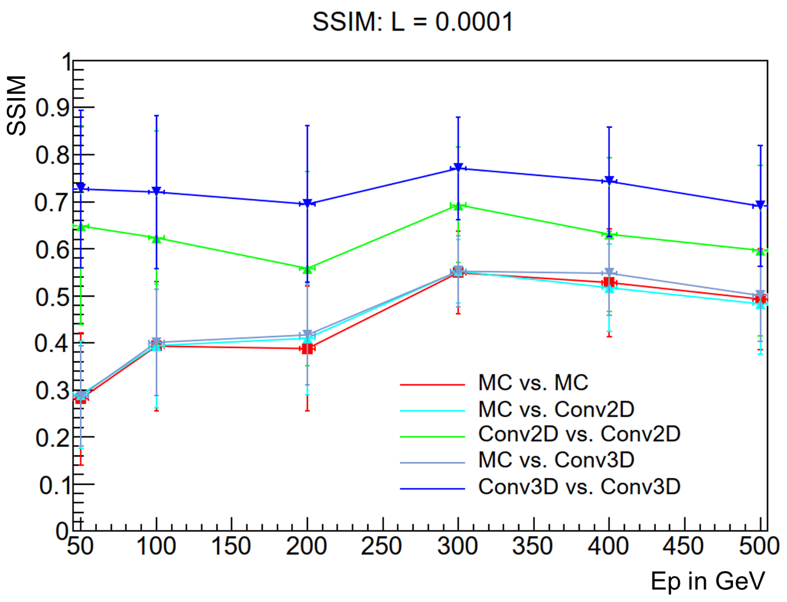}
    \caption{$\mathrm{SSIM}$ between the different image sets.}
    \label{fig:SSIM}
\end{figure}

The original GAN paper \cite{goodfellow} suggests that GANs can learn any target distribution if sufficiently large networks, training samples, and computation time are given. However, the theoretical analysis conducted in \cite{Arora2017} showed that the training objective can approach its optimum value even though the generated distribution is far from the target distribution. Moreover, GAN models suffer from well-known problems, such as mode dropping or mode collapse \cite{mode_dropping}, which affect the quality of the generated sample. Understanding whether GANs can reproduce the same level of similarity exhibited by the original Monte Carlo data set, is essential in the context of scientific simulations.
In order to better understand the level of similarity/diversity among the images, we measure the Structural Similarity Index ($\mathrm{SSIM}$) \cite{SSIM}. The $\mathrm{SSIM}$ estimates the perceptual difference between similar images, a value of $1$ indicating two identical images. The $\mathrm{SSIM}$ is adjusted by the parameter $\mathrm{L}$, which accounts for the pixel intensity dynamic range, and stabilizes for $\mathrm{L}=10^{-4}$. We calculate the $\mathrm{SSIM}$ building 2D windows inside the image and sliding them across the calorimeter $x-y$ plane, while the $z$ axis is treated as the image channel dimension. We measure the index for random image pairs from the Monte Carlo data (labelled as "MC vs MC"), the GAN data ("GAN vs GAN") or both sets ("MC vs GAN"). Figure \ref{fig:SSIM} shows the results, calculated for $\mathrm{L}=10^{-4}$.
Overall, we observe that the $\mathrm{SSIM}$ of generated images vs. generated images of both GAN models is clearly higher than the corresponding Monte Carlo value, indicating that, on average, GAN images are more similar to each other than Monte Carlo images are. Or in other words, that the GANs fail to reproduce the image diversity present in the Monte Carlo sample. The Conv2D model does however perform better with lower $\mathrm{SSIM}$ (green, in figure \ref{fig:SSIM}). It is equally interesting to note that, by measuring the similarity between GAN and Monte Carlo images (light blue and gray) we can conclude that both GAN models produce images that resemble Monte Carlo's with the equivalent level of similarity as the Monte Carlo data set itself.

While we plan to continue our investigation in order to better quantify GAN performance in terms of support space size, sample diversity and mode dropping effects, the current result is well within the precision level required from fast simulation models. The largest discrepancy is observed for single cell energies, shown in figure \ref{fig:flatecal}, which we can adjust by tuning the initial image pre-processing step.

\section{Conclusion and Future Work}
In this paper we introduce a novel Conv2D neural network architecture for simulating high energy physics calorimeter detector outputs. We compare the Conv2D to a prior Conv3D model and validate the results against Monte Carlo simulations. The result evaluation is performed in terms of several physics quantities, such as energy resolution, shower shapes or sampling fraction and image analysis metrics such as the $\mathrm{SSIM}$. All things considered, the Conv2D model achieves a extremely satisfactory level of agreement with respect to Monte Carlo simulations. At the same time, the SSIM index hints to a slightly more limited diversity among the generated images. A full characterisation of the consequences of this difference, with respect to typical simulated data set use cases is needed. Together with a detailed study to understand, whether image similarity can be correlated to any specific bias in the generated data.
On the computational side, despite a larger number of parameters in the Conv2D model, we obtain a significant speed up in terms both the inference and training time. 

\section*{\uppercase{Acknowledgements}}
This work has been sponsored by the Wolfgang Gentner Programme of the German Federal Ministry of Education and Research.

\bibliography{AAAI}

\end{document}